\documentstyle[11pt]{article}

\input{hepsfig.sty}

\begin{document}
\baselineskip=6mm


\begin{center}
{\Large\bf Nonrelativistic phase in $\gamma$-ray burst afterglows }
\footnote{ Supported by the National Natural Science Foundation of 
China (grants numbers 10625313, 10221001, 10473023); 
K. S. Cheng is supported by a RGC grant of the Hong Kong 
Government under HKU 7014/04P.

Email: hyf@nju.edu.cn}

\vspace{5.0mm}
HUANG Yongfeng \\
{\sl Department of Astronomy, Nanjing University, Nanjing 210093}

LU Tan  \\
{\sl Purple Mountain Observatory, Nanjing 210008}

CHENG Kwong Sang \\
{\sl Department of Physics, the University of Hong Kong, Hong Kong}

\end{center}

\vspace{5mm}
\noindent
{\Large\it Accepted for publication in: {\Large\it\bf Progress in Natural Science}} 

\vspace{5mm}
\noindent
{\sl {\bf Abstract: }
The discovery of multiband afterglows definitely shows that most $\gamma$-ray
bursts are of cosmological origin. $\gamma$-ray bursts are found to be 
one of the most violent explosive phenomena in the Universe, in 
which astonishing ultra-relativistic motions are involved. In this article, 
the multiband observational characteristics of $\gamma$-ray bursts and 
their afterglows are briefly reviewed. The standard model of $\gamma$-ray 
bursts, i.e. the fireball model, is described. Emphasis is then put on 
the importance of the nonrelativistic phase of afterglows. The concept of 
deep Newtonian phase is elaborated. A generic dynamical model that is 
applicable in both the relativistic and nonrelativistic phases is introduced.
Based on these elaborations, the overall afterglow behaviors, from the 
very early stages to the very late stages, can be conveniently calculated.  }

\vspace{5mm}

\noindent
{\it Keywords: \hspace{1mm} gamma-ray bursts, afterglows, jets, shock waves, relativity } 
 
\vspace{5mm}

\section{Introduction}

$\gamma$-ray bursts (GRBs) are brief bursts of high energy radiation 
from the sky. They were serendipitously discovered by Vela satellites in 
late 1960s, and were first publicly reported by Klebesadel et al. 
in 1973 $^{\cite{kso73}}$. In their classic paper, Klebesadel et al. 
reported 16 events simultaneously detected by at least two satellites. 
Their duration varies from less than 0.1 s to about 30 s, and the light 
curve can be very complicated. The energies of emitted photons are mainly 
between 0.2 MeV and 1.5 MeV. 

Such strong bursts in $\gamma$-rays are completely unexpected to astronomers, 
thus received wide attention from the beginning. However, the enigma of 
GRBs once puzzled us for a long time, because the observational data 
were so coarse at early times that many basic problems that 
are closely related to observations are uncertain. The most representative 
problem is how far are GRBs from us. Since the localization ability of 
$\gamma$-ray detectors is very poor, astronomers could not find the 
counterparts of GRBs in optical, infrared, and radio bands. It is then 
impossible to measure the distances of GRBs directly.  

In 1980s, many researchers tended to believe that GRBs are within our 
Galaxy. In April 1991, an unprecedentedly sensitive high energy 
satellite, Compton Gamma-Ray Observatory (CGRO), were launched by NASA, leading 
to a rapid increase in the number of observed GRBs. The BATSE instrument 
onboard CGRO unexpectedly found that GRBs are isotropically distributed in 
the sky and they do not show any corelation with the structure of our Galaxy. 
This strongly hints that GRBs are of cosmological origin. A direct measure 
of the distance becomes even pressing. The final breakthrough were made in 
1997: thanks to the successful operation of the X-ray telescope onboard the 
Italian-Dutch BeppoSAX satellite, a few GRBs were rapidly and accurately 
locallized, leading to the discovery of their X-ray, optical and radio 
counterparts (i.e., afterglows), and even their host galaxies. The enigma 
of GRBs began to be unveiled $^{\cite{costa97, van97, frail97}}$. 

Rapid progresses have been made in the field of GRBs since the discovery 
of afterglows. Important results include: the association of long GRBs 
with type Ic supernovae, the observations of early afterglows, and the 
detection of a high redshift GRB. Today, it has been well known that 
GRBs occur at the far end of our Universe. The energy release is so enormous 
that GRBs are believed to be the most violent explosions since the 
Big Bang. The distance problem is resolved, but theorists are now
confronted with an even more tough task of explaining the nature of GRBs. 

A few important review articles are available for GRBs 
$^{\cite{fm95, piran99, vkw00, cl01, m01, m02, zhang04, zhang07}}$. 
In this article, we mainly concentrate on the nonrelativisitc phase 
of GRB afterglows. For completeness, we first introduce the observational 
aspects of GRBs and afterglows in Section 2, and their standard 
theoretical interpretions in Sections 3. Section 4 is devoted to the 
nonrelativistic phase of afterglows. 
Finally, Section 5 is our conclusion and discussion.

\begin{figure}[htb]
  \begin{center}
  \leavevmode
  \centerline{ 
  \epsfig{figure=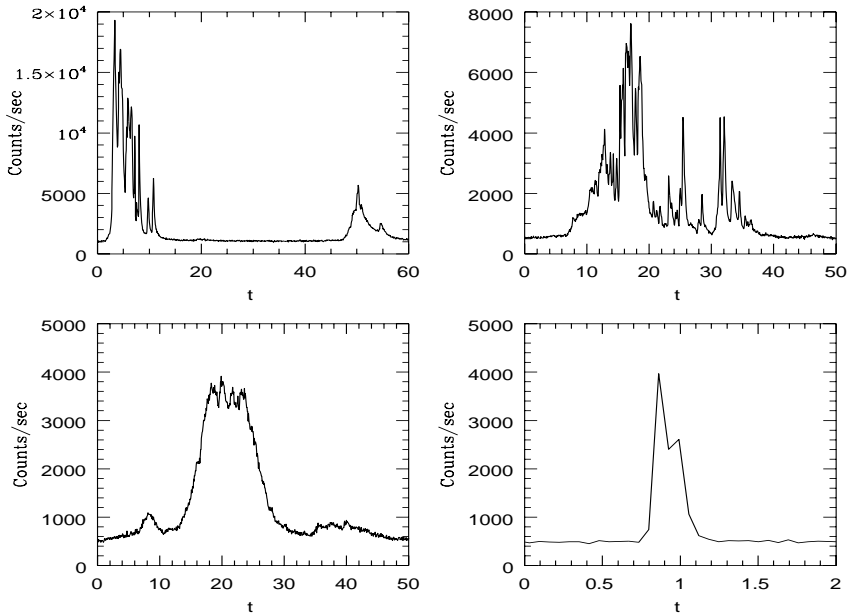,width=3.0in,height=3.0in,angle=0,
  bbllx=150pt, bblly=150pt, bburx=525pt, bbury=650pt}
  }
\caption {Light curves of 4 GRBs observed by BATSE $^{\cite{piran99}}$. }
  \end{center}
  \end{figure}

\begin{figure}[htb]
  \begin{center}
  \leavevmode
  \centerline{ 
  \epsfig{figure=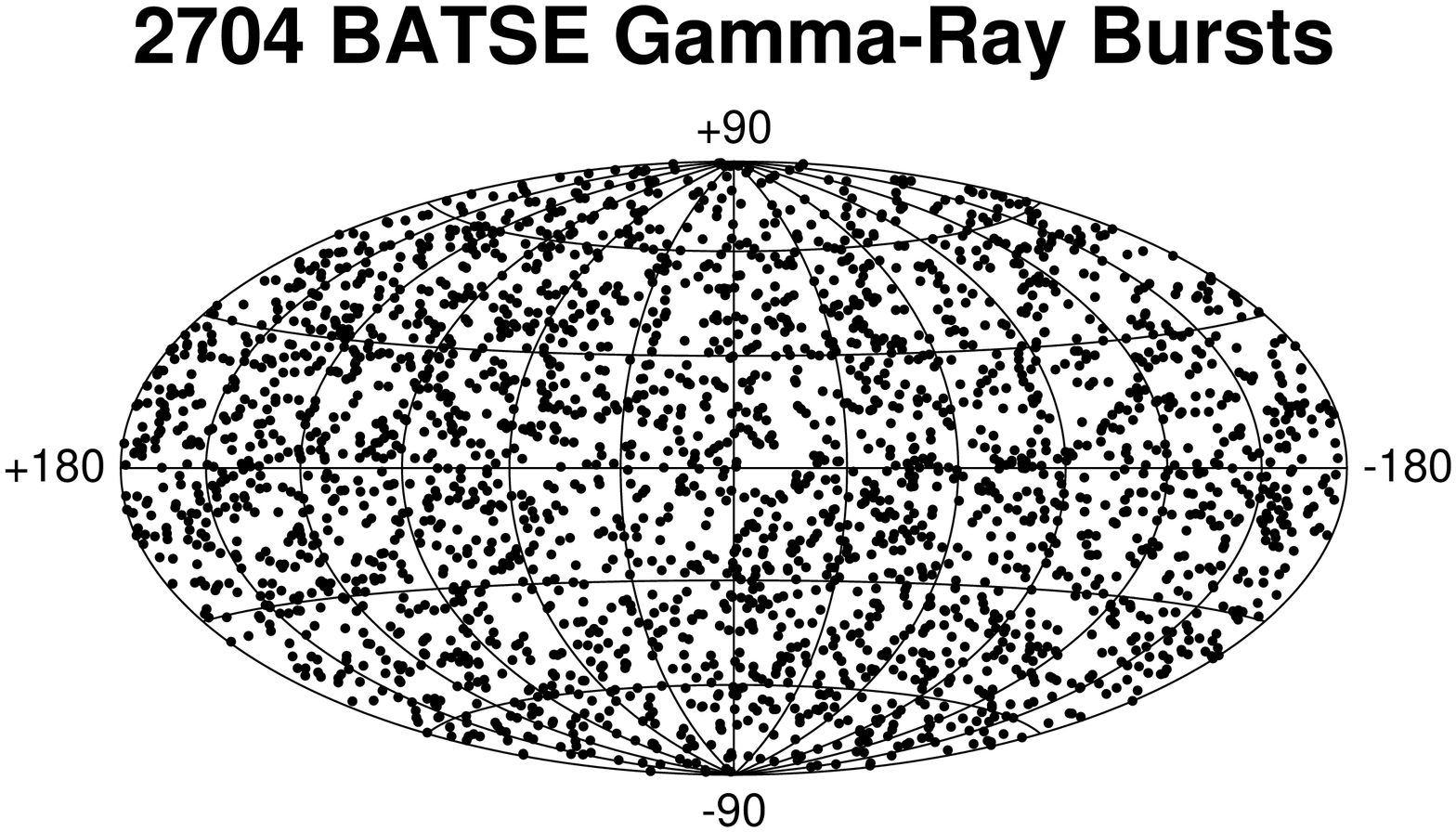,width=3.0in,height=3.0in,angle=90,
  bbllx=100pt, bblly=120pt, bburx=575pt, bbury=580pt}
  }
\caption {The distribution of 2704 GRBs observed by BATSE on the sky 
(http://www.batse.msfc.nasa.gov/batse/grb/skymap/). }
  \end{center}
  \end{figure}

\section{Observational Characteristics} 

GRBs could only be detected shortly in $\gamma$-rays before 1997, which 
seriously prohibited researchers from understanding their nature. Beginning 
from 1997, people find that GRBs are followed by long-lived low energy 
(i.e., X-ray, optical, infrared and radio) afterglows. 
Such observations provide key clues for our 
understanding of these enigmatic objects. Below, we briefly review the 
observational properties of GRBs and their afterglows.    
 
\subsection{GRBs}

The main burst phase of GRBs is usually very short, during which the
emitted photons are mainly $\gamma$-rays. The 
basic features are summarized as follows$^{\cite{fm95}}$:

\begin{itemize}

\item Event rate. During its operation period (1991 --- 2000), BATSE could
detect one or two GRBs almost every day, indicating that they are not 
rare events. Till now, more than 3000 GRBs have been observed, most of which
were contributed from BATSE observations. Since we have known that GRBs occur at cosmological 
distances, it can be easily derived that a GRB will happen in one typical 
galaxy every $10^4$---$10^5$ years (assuming that the radiation is isotropic).

\item Durations. The distribution of GRB durations 
is bimodal, i.e., GRBs can be divided into two sub-groups: long bursts 
with durations clustered at $\sim$ 20 s, and short bursts with durations
clustered at $\sim$ 0.2 s$^{\cite{kou93}}$. Generally speaking, the 
spectra of short bursts are slightly harder than those of long bursts. 
So these two sub-groups are usually called short/hard bursts, and 
long/soft bursts respectively. Long bursts are about three times as 
many as short ones. 

\item Light curves. The temporal structure of GRBs is very complex. The 
light curves of some GRBs are smooth, but in most cases the light curves 
are highly variable. Figure 1 gives some examples. It has been specially 
noted that in a few events, the $\gamma$-ray flux increases rapidly 
from the background to the peak in less than $\delta t \sim 0.2$ ms. 
This provides a strong constraint that the size of the radiation zone 
should be less than $c \delta t \sim 60$ km if the emitter's bulk velocity 
is much less than the speed of light. 

\item Spectral features. The photon energy of a GRB is typically between 
200 keV and several MeV. In some rare cases, it can even extend to $\sim$ 
20 GeV. In the typical energy range, the spectrum is obviously non-thermal, 
and can be well approximated by a power-law or a broken power-law 
function $^{\cite{band93}}$. 

\item Energetics. The typical fluence of GRBs is between $10^{-9}$ --- $10^{-6}$
J/m$^2$. The total radiated energy will be $\sim 10^{45}$ J if the emission 
is isotropic. In some extreme cases, the energy can even be as large as $10^{47}$ 
J, or $\sim 2 M_\odot c^2$. 

\item Sky distribution. BATSE found that the distribution of GRBs is isotropic 
in the sky$^{\cite{mee92}}$, as illustrated in Figure 2.  
No correlation can be found between GRBs and the structure of our Galaxy.
It provides the first clue that GRBs should be of extra-galatic origin. 
Additionally BATSE also found that there are too few weak GRBs
as compared with original expection. It can be interpreted as the effect of 
the expansion of our Universe. Due to these BATSE discoveries, the opinion 
that GRBs should be of cosmological origin became more and more popular in 
the 1990s.  

\end{itemize}

Here we must note that there are actually two kinds of GRBs: soft $\gamma$-ray
repeaters (SGRs) and classical GRBs. SGRs can repeatedly but randomly burst out.  
On the contrary, classical GRBs do not show any repetition activities. 
Only 5 SGR sources have been found till now. More than 100 bursts have been 
observed from the most active SGR, while less than 10 events were observed from 
the most inactive one. Another feature of SGRs is that the average photon 
energy is low, i.e., $\sim 30$ keV. It has been generally accepted that SGRs 
are neutron stars with super-strong magnetic field (i.e., magnetars) in local 
galaxies$^{\cite{woods}}$. In this article, all ``GRBs'' are referred to 
classical GRBs unless declared explicitly otherwise.

\begin{figure}[htb]
  \begin{center}
  \leavevmode
  \centerline{ 
  \epsfig{figure=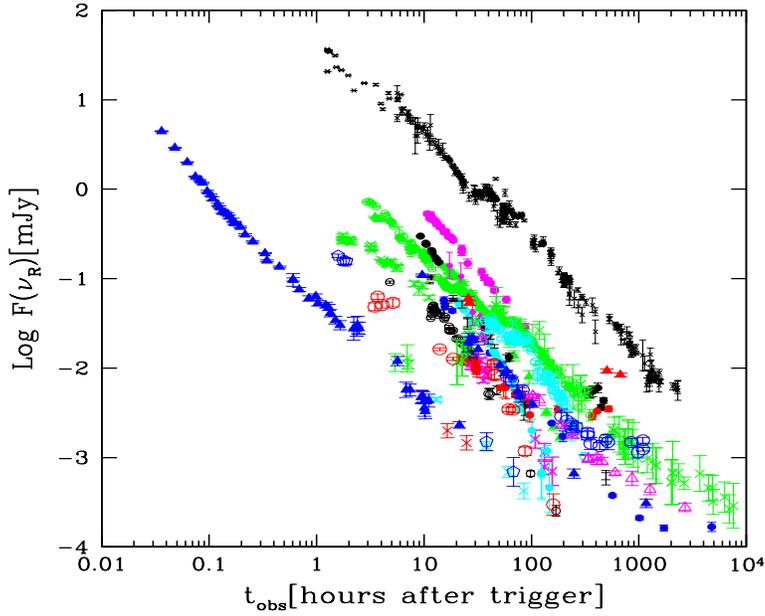,width=3.0in,height=3.0in,angle=0,
  bbllx=150pt, bblly=200pt, bburx=525pt, bbury=650pt}
  }
\caption {R band afterglow light curves of 24 GRBs$^{\cite{nard06}}$. }
  \end{center}
  \end{figure}

\begin{figure}[htb]
  \begin{center}
  \leavevmode
  \centerline{ 
  \epsfig{figure=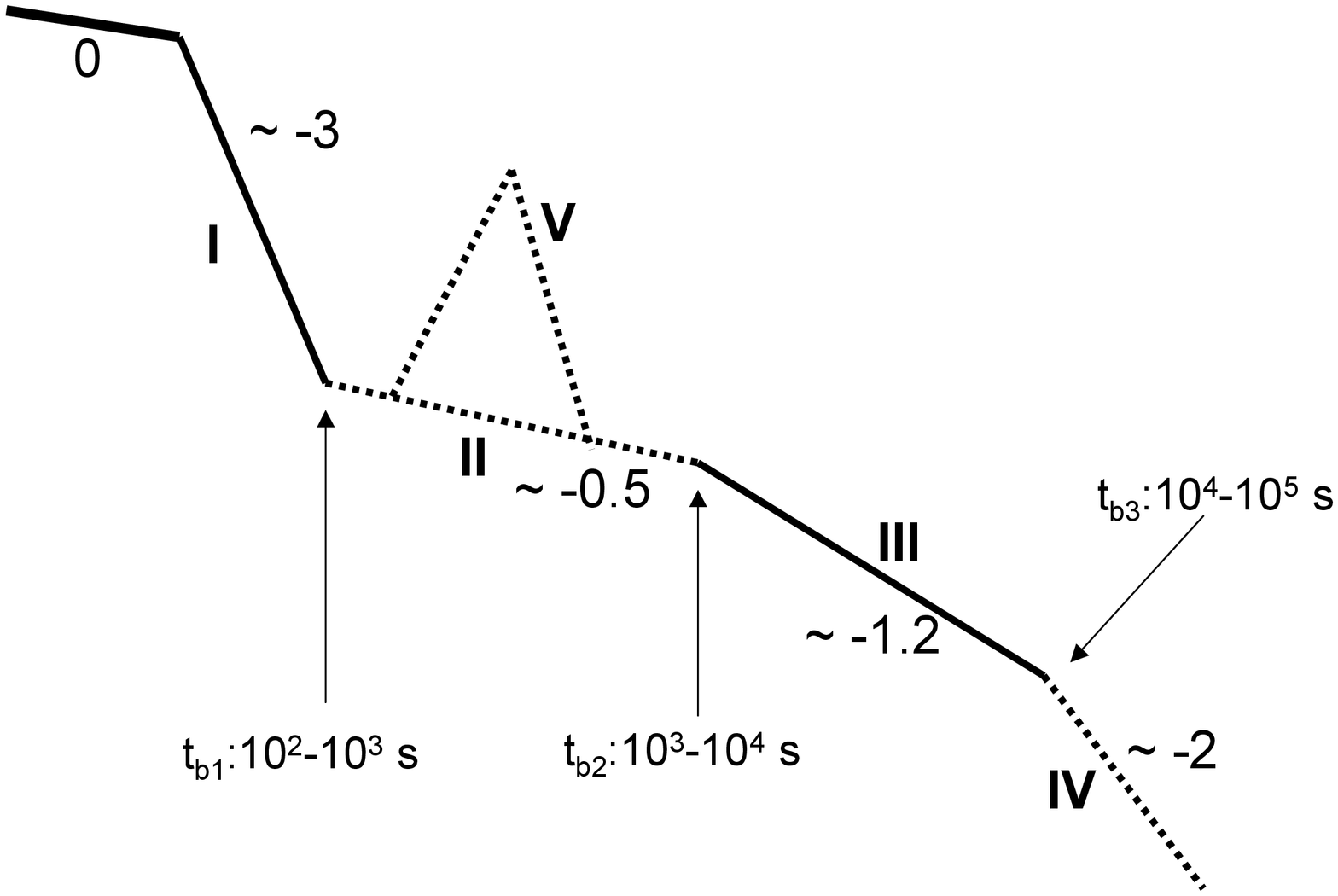,width=3.0in,height=3.0in,angle=270,
  bbllx=150pt, bblly=200pt, bburx=525pt, bbury=550pt}
  }
\caption {A Cartoon illustration of the X-ray afterglow
of GRBs $^{\cite{zhangfd06}}$. }
  \end{center}
  \end{figure}

\subsection{Afterglows}

It has long been expected that GRBs should be associated with afterglows. 
However, due to the awkward localization ability, afterglow was  
detected for the first time only till 1997. 
In the operation period of BeppoSAX (1997 --- 2000), only one or two GRBs
could be localized each month. On Dec 20, 2004, the Swift satellite 
was launched$^{\cite{geh04}}$. It can localize nearly 200 GRBs every year,  
leading to a rapid increase in the afterglow sample. Till the end of January 
2007, $\sim 480$ GRBs have been rapidly localized, of which $\sim$80\% are 
detected in X-rays, $\sim 40$\%(i.e. 176 events) are detected in optical or 
infrared bands, and $\sim 10$\% are detected in radio. Redshifts have been 
measured for $\sim 20$\% (111) GRBs, with the highest record being 
$z=6.29$ for GRB 050904$^{\cite{cus06, hai06, kaw06}}$. 

Before 2005, afterglows were usually detected several hours after the
main bursts due to the limited localization ability. At this stage, 
the optical/infrared afterglow generally decays as a power-law function of time, 
$S_{\rm \nu} \propto t^{-1.0}$ --- $t^{-1.4}$. Several days later, 
the decay becomes steeper as $S_{\rm \nu} \propto t^{-2.0}$ --- $t^{-2.5}$, 
so that a break can be observed in the afterglow light 
curve$^{\cite{kul99, har99}}$. This break can be well explained by 
the jet effect, which will be further illustrated later. The optical/infrared 
afterglow light curves are generally smooth (see Fig. 3 for 
some examples$^{\cite{nard06}}$), although rebrightening can 
also be occasionally seen in a few events. 
In the radio bands, afterglow behaviors are some what different. 
When the observer's time ($t$) is less than 20 --- 40 days, the radio 
emission is in a brightening phase. Fast variability can also be 
observed, which should be scintillation due to the scatter by the 
interstellar medium. When $t >20$ --- 40 days, the radio afterglow 
decays as a power-law of time, and the scintillation also 
disappears$^{\cite{frail97}}$. 

Thanks to the quick response of Swift satellite, very early afterglows are 
observed from many GRBs since 2005. In a few cases, afterglows are 
observed even for $t < 20$ s. At the very early stage ($t \leq $ 100 --- 
300 s), the X-ray afterglow decays steeply, then it enters a shallow
decay phase that may last for a few hours, and finally connects with 
the later afterglow as described in the above paragraph$^{\cite{obr06}}$.
An unexpected result of Swift is that X-ray flares were detected 
during the period of 100 s  $\leq t  \leq 10000$ s in many 
GRBs$^{\cite{bur05, chin07}}$. These enigmatic flares can hopefully 
provide useful clues for our understanding of the central engine of GRBs. 

The X-ray afterglows of GRBs show some general 
characterics$^{\cite{zhangfd06, nous06}}$, as illustrated in Fig.~4.
In this cartoon figure, the segment marked with ``0'' indicates emission
of the main burst phase; Segment ``I'' indicats the early fast decay phase,
usually with $S_{\rm \nu} \propto t^{-3}$ --- $t^{-5}$; Segment ``II'' 
indicates the subsequent shallow decay phase, with 
$S_{\rm \nu} \propto t^{-0.2}$ --- $t^{-0.8}$; Segment ``III'' is the 
late normal decay phase, $S_{\rm \nu} \propto t^{-1.0}$ --- $t^{-1.3}$; 
Segment ``IV'' is the late fast decay phase, 
$S_{\rm \nu} \propto t^{-2.0}$ --- $t^{-2.5}$; The dashed segment ``V'' 
refers to possible flares at early stage. Note that multiple flares may 
be observed in a single event. It should be pointed out that not every 
GRB has all the five segments. Actually most GRBs lack one or two of 
these components, and very few events can have all the five components. 

\section{Fireball Model}

The discovery of afterglows leads to a final solution for the 
famous distance problem that once troubled astronomers for about 30 years. 
The enigma of GRBs is now being unveiled. We now know that in a typical 
GRB event, a huge amount of energy ($10^{43}$ --- $10^{47}$ J) will be 
released from a small volume (on the scale of tens to hundreds 
of kilometers) in a very short period (tens of seconds). This will 
inevitably give birth to a fireball whose optical depth will be much 
larger than 1. Emission from the surface of the fireball cannot 
account for the observed GRB, since the luminosity will be too low and 
also its thermal spectrum is not consistent with the observed non-thermal
spectrum. However, we can imagine that the radiation pressure inside the 
fireball should be enormous, which will drive the fireball to expand 
outward. If the fireball is mainly composed of electron-positron pairs 
and photons so that baryon pollution is small (with a static 
mass less than $10^{-7}$ --- $10^{-3} M_\odot$),  then the fireball 
material will be accelerated to an ultra-relativistic speed at the radius of 
$R \sim 10^5$ --- $10^8$ km, forming a thin shell with the bulk Lorentz 
factor $\gamma >$ 100 --- 1000. At $R\sim 10^{11}$ km, the shell will be
decelerated by interstellar medium, producing strong blastwaves. These 
blastwaves are called external shocks since they originate from the 
interaction of the fireball material with the external medium. Electrons 
are accelerated by shocks, whose emission may give birth to the 
observed GRB. 

However, the large radii of the external shocks indicate that they cannot
produce a highly variable light curve as observed in the main burst 
phase. It has been suggested that the central engine may be active 
for tens of seconds, producing many thin shells with different 
velocities, like a geyser. These shells will collide with each other 
at $R \sim 10^8$ km and produce a sequence of shocks, giving birth to 
a highly variable GRB. These shocks are called internal shocks since 
they come from the interation of the material within the fireball. Now 
the most popular view is that the main bursts are produced by 
internal shocks, and afterglows are produced by long-lasting external 
shocks. This is the so called standard fireball 
model$^{\cite{pacz86, good86, shem90, rees92, mes93, rees94, pacz94}}$. 

The Fireball model can explain the observed afterglows well. For 
example, in Fig.~4, the fast decay of Segment I comes from 
the delayed high latitude emission when the internal shocks quenched; 
The shallow decay of Segment II may be due to continuous energy 
injection from the central engine; The normal decay of Segment III 
can be naturally explained by considering the deceleration of the
external shock; The fast decay of Segment IV indicates that the 
GRB ejecta is not isotropic, but should be a jet with a typical 
half-opening angle of $\sim 0.1$; Finally, the flares of Segment V 
can be interpreted as late explosions of the central engine. Additionally, 
the giant amplitude scintillation of radio afterglows at early stage 
can be naturally regarded as a proof for ultra-relativistic motion. 
The radius of the fireball is relatively 
small at first, so that the scattering of interstellar medium is 
serious. But tens of days later, the fireball 
has increased markedly due to ultra-relativistic expansion, then 
scintillation becomes insignificantly. 

The above explanation of afterglows is generally satisfactory and 
the physics of afterglows are relatively clear. However, we still 
know very few about the central engine of GRBs. After the initial 
acceleration phase, the fireball loses most of its memory to the 
progenitor and we can hardly find any direct clues from afterglows. 
From the fireball model, we only know that the central engine 
must satisfy the following requirements:  (i) Can release a huge 
amount of radiation energy (typically $\sim 10^{45}$ J, but can also 
as large as $10^{47}$ J in some extreme cases); (ii) The energy 
is released from a small volume (less than 100 km); (iii) The energy 
release process should last for several seconds to tens of seconds, and 
should be intermittent; (iv) The fireball should be free of baryon 
pollution, i.e., the static mass of baryons associated with every 
$\sim 10^{45}$ J of energy should be less than $\sim 10^{-5} 
M_\odot$; (v) The event rate should be one GRB per typical galaxy
every $\sim 10^6$ --- $10^7$ year. The above discussion is based on 
the assumption that the radiation of GRB is isotropic. If GRB 
emission is highly collimated, then some requirements will be 
changed. For example, the intrinsic energy may be lower by two 
magnitudes, but the event rate per galaxy should be correspondingly 
higher. 

In any case, the above requirements are rigorous and can be satisfied 
only by very few objects. Currently there are mainly two kinds of models: 
(i) Merge of compact stars, including the merge of two neutron stars, or 
a neutron star with a black hole; (ii) Collapse to a black hole of a 
dying massive star ($M \geq 40 M_\odot$). A similar structure will be 
formed in these two processes, a quickly rotating black hole (or neutron 
star) surrounded by an accretion disk, plus two highly collimated jets
moving outward perpendicular to the disk. In Model (i), the lifetimes of the 
accretion disk and jets are very short, much less than 1 second. It 
is suitable for producing short GRBs. On the contrary, the lifetimes
of disk and jets in Model (ii) can be tens of seconds and it is 
proper for long GRBs. Now it has been widely believed that long GRBs 
should be due to massive star collapses and short GRBs be due to 
mergers of compact stars. This viewpoint is further supported by 
the observational facts that long GRBs usually occur in star forming
regions and are frequently associated with type Ic supernovae, while 
short GRBs generally deviate from star forming regions. 

However, there are still many unsolved problems. For example, why 
the central engine behaves like a geyser, as required by the internal 
shocks? How can the central engine avoid the baryon pollution problem, 
so that the fireball material can be accelerated to ultra-relativistic 
speeds? How are highly collimated jets produced? How can the central 
engine be active for hundreds or even thousands of seconds, so as 
to produce the observed flares (Segment V in Fig.~4) and continuous 
energy enjection (Segment II in Fig.~4)? Why X-ray and 
optical afterglows behave quite differently in some GRBs? 
Why there are many dark GRBs, i.e., GRBs with afterglows detected in X-rays
but not in optical bands? What is the true relation between GRBs and 
supernovae? Additionally, the connections between GRBs and cosmology, cosmic
rays, gravitational waves are all attracting problems and need many 
hard work. 

\section{Afterglows in the Nonrelativistic Phase}

The fireball model is most successful in explaining Segments III and 
IV in Fig.~4, which are all due to external shocks. According to the 
standard fireball model, the external shock should be ultra-relativistic 
(Lorentz factor $\gamma \gg 1$) initially. In a homogeneous external medium, it 
decelerates as $\gamma \propto t^{-3/8}$, $R \propto t^{1/4}$, $\gamma \propto 
R^{-3/2}$ in the adiabatic case. Then it is easy to derive that synchrotron 
radiation from the external shock should decay as a power-law of time. For 
a highly collimated jet with a half opening angle $\theta$, 
lateral expansion will be significant when $\gamma < 1/\theta$,
leading to a more quick deceleration. At the same time,
the edge of the jet becomes visible. These two factors act together to make 
the afterglow decay more steeply. This is the reason for the observed light curve 
break (the so called ``jet break'') between Segments III and IV$^{\cite{rhoads97}}$.  
  
As mentioned above, a basic assumption of the standard fireball model is that 
the external shocks are ultra-relativistic ($\gamma \gg 1$). However, we 
will point out below that this assumption is correct only in a very limited period.

\subsection{Importance of the Nonrelativistic Phase}

GRBs are most impressive for two characteristics: the huge energy release and 
the ultra-relativistic motion. Misled by the energetics, people once believe 
that the fireball should be ultra-relativistic for a long time in the 
afterglow phase. As such, the explanation of the fireball model to afterglows is 
based on the assumption of $\gamma \gg 1$. However, a careful study of the 
deceleration of the external shock reveals that it is not true. The fireball 
generally decelerates as $\gamma \propto t^{-3/8}$, with the detailed 
expression given by, 
\begin{equation}
   \gamma \approx (200 - 400) E_{44}^{1/8} n_0^{-1/8} t_{\rm s}^{-3/8}, 
\end{equation}
where $E_{44}$ is isotropic kinetic energy in units of $10^{44}$ J, 
$n_0$ is the number density of the interstellar medium in units of 1 cm$^{-3}$, 
and $t_{\rm s}$ is observer's time in units of second. In typical cases, 
we would find that $\gamma \approx$ 2.8 --- 5.6 for $t=1$ day, 
$\gamma \approx$ 1.2 --- 2.4 for $t=10$ day. For $t=30$ day, the Lorentz 
factor will even be less than 1, which is of course un-physical. 
In fact, Eq. (1) is an approximate expression that is applicable  
only when $\gamma \gg 1$. In realistic case, the condition of 
$\gamma \gg 1$ may no longer be satisfied several days or teens of
days later, then Eq.~(1) cannot be used to calculate the afterglows. 

Observationally, X-ray afterglows usually last for one or two weeks, 
optical afterglows can last for tens of days to several months, and 
radio afterglows can even last for more than 1000 
days$^{\cite{frail03, frail04, horst}}$. To account for these observations, 
we will inevitably need to consider the dynamics and radiation in 
the nonrelativistic phase$^{\cite{berger04}}$. In fact, 
the importance of nonrelativistic phase has been pointed out by 
Huang et al. as early as in 1998, i.e., soon after the discovery of 
GRB afterglows$^{\cite{huang98aa, huang98mn}}$.

\subsection{Generic Dynamical Model}

Now we describe the overall evolution of the external shocks. At first, 
the shock is ultra-relativistic, with $\gamma > 100$ --- 1000. The 
shock is also highly radiative since the synchrotron-induced energy 
loss rate ($P_{\rm syn}$) is much higher than the rate due to 
adiabatic expansion ($P_{\rm ex}$), leading to a high radiation efficiency 
of $\epsilon \sim P_{\rm syn}/(P_{\rm syn}+P_{\rm ex})\approx 1$. The 
deceleration of such a highly radiative ultra-relativistic shock in a 
homogeneous external medium is $\gamma \propto t^{-3/7}$, $R \propto t^{1/7}$, 
$\gamma \propto R^{-3}$. With the increase of radius, the ratio of 
$P_{\rm syn}$ to $P_{\rm ex}$ decreases. Several hours later, $\epsilon$
will be very close to 0 so that the shock becomes adiabatic, but it is 
still ultra-relativistic ($\gamma \gg 1$). The evolution laws then become
$\gamma \propto t^{-3/8}$, $R \propto t^{1/4}$, $\gamma \propto R^{-3/2}$. 
Several days later, $\gamma$ will be decreased to be between 2 and 5. 
At this stage, the approximate equations under $\gamma \gg 1$ assumption
may still be marginally applicable. But teens or tens of days later, 
$\gamma$ will be so close to 1 that the shock will no longer be highly 
relativistic. It enters the nonrelativistic phase. According to the well-known
Sedov solution$^{\cite{sedov}}$, the evoluton of such a nonrelativistic adiabatic 
shock should be $v \propto t^{-3/5}$, $R \propto t^{2/5}$, $v \propto R^{-3/2}$, 
where $v$ is the velocity of the shock. 

From the above description, we see that there are two transitions in the 
process. One is the transition from highly radiative regime to adiabatic regime 
at about a few hours, the other is the transition from relativistic phase to 
nonrelativistic phase at about teens of days. The overall evolution can then 
be divided into three stages: (i) ultra-relativistic and highly radiative
stage ($t \leq$ a few hours), $\gamma \gg 1, \epsilon \approx 1$; 
(ii) ultra-relativistic and adiabatic stage (a few hours $< t <$ a few days), 
$\gamma \gg 1, \epsilon \approx 0$; (iii) nonrelativistic and adiabatic 
stage ($t \geq$ teens of days), $v \ll c, \gamma \approx 1, \epsilon \approx 0$. 
To calculate the afterglows, a conventional method is to use approximate 
expressions within each stage separately. The overall light curve can then be 
obtained by connecting the three segments. However, a natural problem in the 
process is that the light curve may be discontinuous at the 
joint points. Additionally, if the GRB ejecta is a jet, we must go further to 
consider the effect of the lateral expansion. This will lead to a fourth segment. 
If the cooling-induced broken power-law distribution of 
electrons is taken into account, then many additional segments will be 
added. What makes things even more complicated is that the order of all these 
segments depends on parameters. Finally, the analytical expressions for the 
overall light curve become very complicated$^{\cite{zhang04}}$.

\begin{figure}[htb]
  \begin{center}
  \leavevmode
  \centerline{ 
  \epsfig{figure=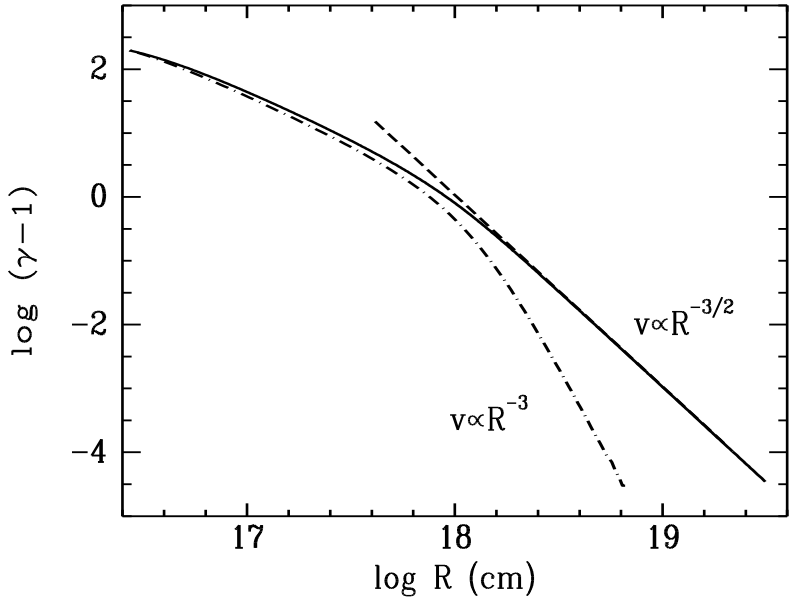,width=3.0in,height=3.0in,angle=270,
  bbllx=150pt, bblly=200pt, bburx=525pt, bbury=550pt}
  }
\caption {Lorentz factor versus radius for an isotropic fireball. 
The dashed line corresponds to the Sedov limit, the dash-dotted line is plot 
according to an earlier dynamical model suggested by others, which deviates from 
the Sedov limit markedly. The solid line is plotted according to 
Eq.~(\ref{dgdm}), which is correct in both the relativistic and nonrelativistic 
phases$^{\cite{huang00}}$.  }
  \end{center}
  \end{figure}

To relax the problem, Huang et al. proposed a generic dynamical model
to depict the overall evolution of the external shock$^{\cite{huang99mn}}$,
\begin{equation}
\label{dgdm}
\frac{d \gamma}{d m} = - \frac{\gamma^2 - 1}
       {M_{\rm ej} + \epsilon m + 2 ( 1 - \epsilon) \gamma m}, 
\end{equation}
where $m$ is the mass of the swept-up medium and $M_{\rm ej}$ is the initial mass 
of the GRB ejecta. Taking $\epsilon =1$, Eq.~(\ref{dgdm}) describes a highly 
radiative shock; while taking $\epsilon=0$, Eq.~(\ref{dgdm}) describes an adiabatic 
shock. This equation is applicable in both the ultra-relativistic and nonrelativistic 
phases. 

Eq. (\ref{dgdm}), together with the following equations about the evolution of
swept-up mass, shock radius, and jet opening angle, can 
give a thorough description of the dynamics of external shocks$^{\cite{huang99mn, 
huang00apj, huang00mn}}$, 
\begin{equation}
\label{dmdr2}
\frac{d m}{d R} = 2 \pi R^2 (1 - \cos \theta) n m_{\rm p},
\end{equation}
\begin{equation}
\label{drdt1}
\frac{d R}{d t} = \beta c \gamma (\gamma + \sqrt{\gamma^2 - 1}),
\end{equation}
\begin{equation}
\label{dthdt3}
\frac{d \theta}{d t} 
             = \frac{c_{\rm s} (\gamma + \sqrt{\gamma^2 - 1})}{R},
\end{equation}
where $\beta = v/c$, $m_{\rm p}$ is the proton mass, and $c_{\rm s}$ is 
the sound speed in co-moving frame. In Fig.~5, we plot the Lorentz-radius 
relation calculated according to Eq. (\ref{dgdm}---\ref{dthdt3}). It can 
be seen that Eq. (\ref{dgdm}) is really correct in both relativistic and 
nonrelativistic phases. 

\subsection{Deep Newtonian Phase}

Afterglow mainly comes from the synchrotron radiation from shock-accelerated 
electrons, although inverse Compton scatering may also play a role 
in some particular cases. Now we concentrate on the calculation of 
radiation. The acceleration of electrons is a complicated process. To 
simplify the problem, we usually assume that the accelerated electrons obey 
a power-law distribution according to their energies$^{\cite{light79, you83}}$,
$dN_{\rm e}' / d\gamma_{\rm e} \propto \gamma_{\rm e}^{-p}, 
(\gamma_{\rm e,min}\leq\gamma_{\rm e}\leq\gamma_{\rm e,max})$. 
When the cooling effect is taken into account, the power-law function
will be broken into several segments$^{\cite{sari98}}$. Here, $p$ is 
the power-law index, which is typically between 2 and 3; $\gamma_{\rm e,max}$
is the maximum Lorentz factor of electrons, depending on the equilibrium 
between the acceleration and radiation power; $\gamma_{\rm e,min}$ 
is the minimum Lorentz factor, which actually is also a measure of the 
typical Lorentz factor since very few electrons are accelerated 
to high energy state and most electrons have relatively lower 
energies. The exact expression for $\gamma_{\rm e,min}$ is difficult 
to derive from the acceleration process. Usually we assume an energy 
equipartition between electrons and protons, then we have, 
$\gamma_{\rm e,min} = \xi_{\rm e} (\gamma - 1) 
             m_{\rm p} (p - 2) / [m_{\rm e} (p - 1)] + 1$, 
where $\xi_{\rm e}$ is the equipartition factor. Note that the 
distribution functions discussed here are applicable only to 
ultra-relativistic electrons. 

The generic dynamical equation (\ref{dgdm}) is applicable even 
in the very late phase. But to calculate the afterglow radiation,
we must pay special attention to a problem that is related to 
$\gamma_{\rm e,min}$. According to the energy equipartition 
between electrons and protons, $\gamma_{\rm e,min}$ could be much larger 
than 1 even at the nonrelativistic phase when $\gamma = 1.01$. However, 
with the further deceleration of the external shock, $\gamma_{\rm e,min}$
will inevitably approach 1, so that the electron distribution function 
introduced in the above paragraph will no longer be applicable. 
This stage is called the deep Newtonian phase by Huang \& 
Cheng$^{\cite{huang03mn}}$. Afterglows may enter the deep Newtonian 
phase when $t > $ a few months (see Fig.~6). This factor should 
be taken into account when explaining the observed afterglows that 
can last for several months or even several 
years$^{\cite{frail03, frail04, horst, berger04}}$. 

Huang \& Cheng suggested that in the deep Newtonian phase, 
the basic form of the electron distribution function should 
be revised as$^{\cite{huang03mn}}$£¬
\begin{equation}
\label{eq8}
\frac{d N_{\rm e}'}{d \gamma_{\rm e}} \propto (\gamma_{\rm e} - 1)^{-p} , 
\;\;\; (\gamma_{\rm e,min} \leq \gamma_{\rm e} \leq \gamma_{\rm e,max}).
\end{equation}
In this equation, electrons are still distributed according to 
their energies, but the originally approximate expression for energy 
(i.e. $\gamma_{\rm e}$) is now replaced by the exact expression 
of $\gamma_{\rm e} - 1$. In the deep Newtonian phase, most electrons 
are nonrelativistic, but a minor portion of electrons 
are still relativistic according to Eq.~(\ref{eq8}). Synchrotron 
radiation from these relativistic electrons produces the afterglow 
in the deep Newtonian phase. 

Fig.~6 illustrates the theoretical afterglow light curves
based on the above considerations$^{\cite{huang03mn}}$. 
We see that for an isotropic 
fireball, the afterglow decays slightly more rapidly after entering 
the deep Newtonian phase. In the jet cases, an obvious light curve 
break can be seen when the afterglow transits from ultra-relativistic 
phase to midly-relativistic phase. It is actually due to the jet 
effect, but note that the break usually occurs in the trans-relativistic
phase (when $\gamma \sim$ 2 --- 5) $^{\cite{huang00apj, huang00mn, huang00}}$. 
In the deep Newtonian phase, the light curve becomes slightly
shallower$^{\cite{huang03mn}}$. These results are consistent with 
others' analytical solutions$^{\cite{wdm, livio00}}$.

\begin{figure}[htb]
  \begin{center}
  \leavevmode
  \centerline{ 
  \epsfig{figure=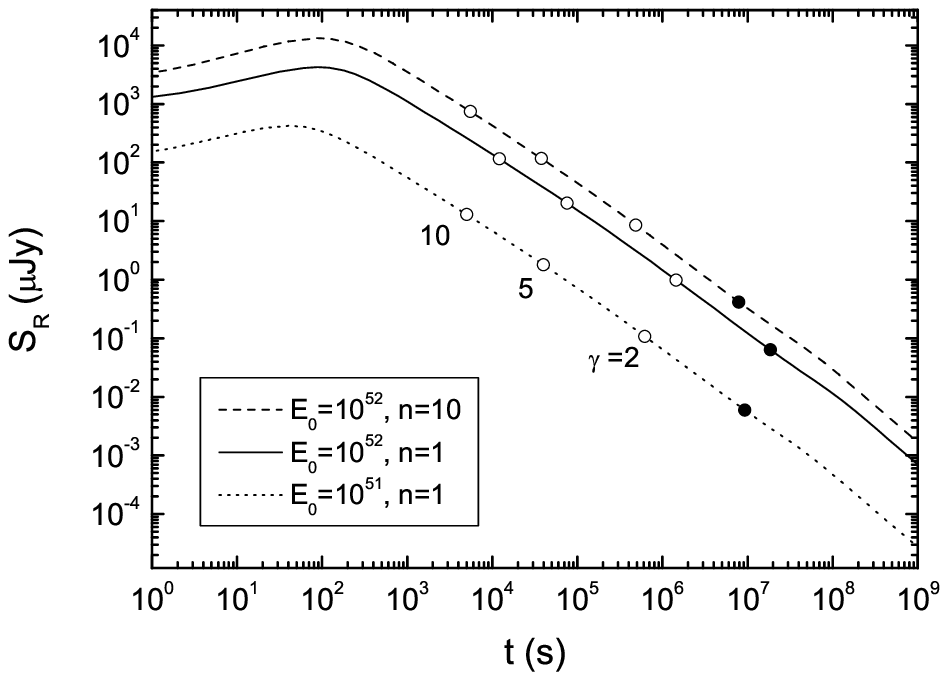,width=3.0in,height=3.0in,angle=0,
  bbllx=20pt, bblly=0pt, bburx=300pt, bbury=250pt}
  \epsfig{figure=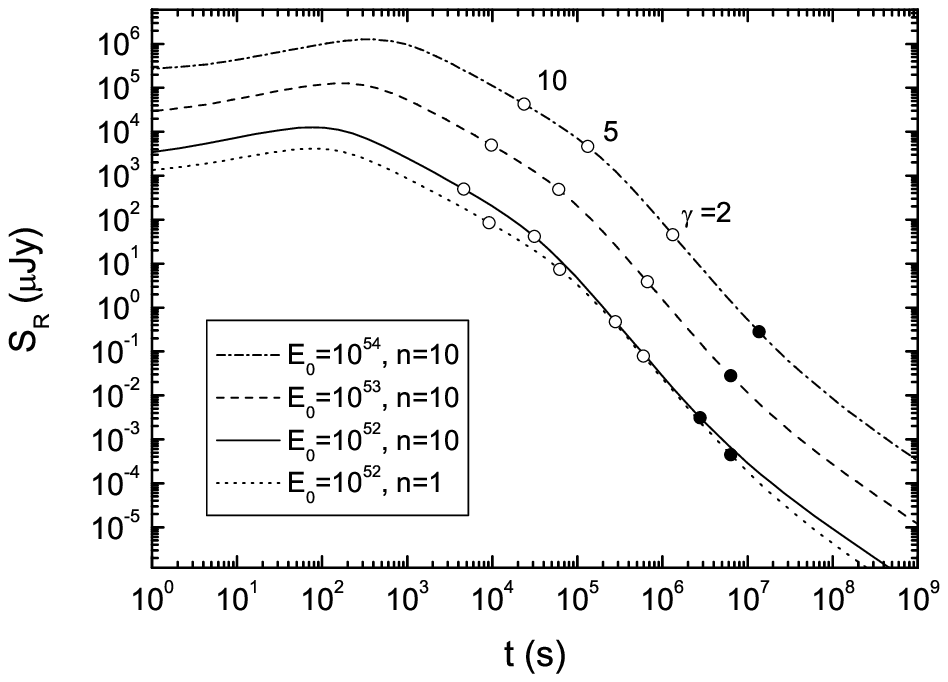,width=3.0in,height=3.0in,angle=0,
  bbllx=20pt, bblly=0pt, bburx=300pt, bbury=250pt}
  }
\caption {The overall afterglow light curve that 
includes the deep Newtonian phase$^{\cite{huang03mn}}$. The left panel 
is plot for isotropic fireballs, and the right panel is for 
jets. Different line styles correspond to various energy (in units 
of ergs) and circum-burst medium density (in units of $cm^{-3}$). 
The decrease of the bulk Lorentz factor $\gamma$ is marked by the 
open circles, and the full circles indicate the time when the afterglow 
enters the deep Newtonian phase. }
  \end{center}
  \end{figure}

\section{Conclusions and Discussion}

The discovery of afterglows is a major breakthrough in the GRB field. 
It definitely 
tells us that at least most GRBs are of cosmological origin. A huge
amount of energy is released in each event, and ultra-relativistic
motions are involved in the process. The fireball model becomes the 
standard model, which can explain the basic features of GRBs and
afterglows well. Here we have given a brief description of the 
observations and theories of GRBs and their afterglows. 
We pay special attention on the importance of the nonrelativistic 
phase in afterglows. A generic dynamical model for the afterglows
is descripted, and the concept of the deep Newtonian phase is 
highlighted. 

Typically the fireball enters the nonrelativistic phase in several 
days or teens of days, and may enter the deep Newtonian phase in 
tens of days or several months. When the GRB ejecta is highly 
collimated, the nonrelativistic phase and deep Newtonian phase 
will come even earlier$^{\cite{huang00apj, huang00mn}}$. 
In some special cases such as when the medium 
is very dense$^{\cite{dailu99, daiwu03}}$, the fireball 
may even enter the nonrelativistic phase in one day. 
In some GRB explosions, the fireball may fail to be accelerated 
to $\gamma \sim 100$ --- 1000 since the energy release is not 
enough or the baryon contamination is too serious. Such a fireball 
with $1 \ll \gamma \ll 100$ cannot produce a GRB successfully, 
but can only give birth to a soft $\gamma$-ray burst or an 
X-ray transient. These events are called failed GRBs by 
Huang et al.$^{\cite{huang02}}$. In these cases, the fireball 
will also enter the nonrelativistic and deep Newtonian phases 
earlier, since the initial Lorentz factor itself is small
and the medium density may be high as well. 

It is interesting that external shocks similar to those 
in GRB afterglows may also exist 
in other explosive phenomena in the cosmos. For example, 
transient X-ray bursts can be accidently observed from the 
center of normal galaxies$^{\cite{halpern04}}$, which may 
be due to the tidal disruption of a star by the massive 
black hole that resides at the galaxy center. Radiation 
is generally believed to come from a temporary accretion 
disk formed in the process. However, Wong et al. suggested that 
external shocks should also be excitated when the jet 
associated with the accretion disk interacts with external 
medium. They found that emission from these external shocks 
can explain the observed X-ray light curves and spectra 
well$^{\cite{wong}}$. In these processes, the external 
shock will also enter the nonrelativistic and deep Newtonian 
phases quickly since the initial lorentz factor of the jet 
is rather low (typically $\gamma < 10$). 

The study of GRBs is a rapidly developing field. In the past 
BeppoSAX era (1997 --- 2002), afterglows were discovered from 
long GRBs for the first time. In the current Swift era 
(beginning from 2004), many encouraging results are further 
brought about, such as the discovery of afterglows from 
short GRBs, the discovery of flares in early afterglows, 
and the detection of the most distant GRB up to a redshift 
of $z \sim 6.3$ $^{\cite{zhang07}}$. However, the enigma of GRBs is far 
from clear yet. Many key problems are still remained uncertain, 
as pointed out in Section 3. For example, a suspicious problem
is related to the beaming of GRBs. In the BeppoSAX era, jet 
breaks have been clearly observed in the afterglow light curves of 
many GRBs. But in the Swift era, when our sample is greatly increased,
very few GRBs show such a jet break in their afterglow light curves.
Are GRBs really highly collimated? An once resolved problem now seems
to be raised again. It is interesting to note that the study of 
nonrelativistic afterglows may help to solve this puzzle, since
the emission in this late phase should be largely isotropic, thus 
it is easier to derive the intrinsic kinetic energy of the GRB 
remnant. A comparison of this intrinsic kinetic energy with the 
prompt gamma-ray energy release then may hopefully give a direct measure 
of the beaming angle. To do this, we need more events with well 
sampled afterglow data that extend to months after the GRB trigger. 
It is encouraging that a powerfull satellite, the Gamma ray Large Area 
Space Telescope (GLAST), will be launched in early 2008. It will
open a new era for the GRB study in the sense that a high energy 
window (20 --- 300 GeV) of GRBs will be open$^{\cite{glast}}$, and more 
afterglow samples will be available as well. We expect that the enigma 
of GRBs will be further revealed in the coming GLAST era.

\end{document}